\documentclass{article}

\usepackage{natbib}
\usepackage{kashsty}
\usepackage{booktabs}
\usepackage[margin=1in]{geometry}

\usepackage{color}

\def\PICDIR{.}

\newcommand{\mc}[1]{\mathcal{#1}}

\title{
 Markov models for ocular fixation locations in the presence
 and absence of colour  
}

\author{
  Adam B Kashlak,
  Eoin Devane, Helge Dietert, and Henry Jackson 
}

\begin{document}

\maketitle

\begin{abstract}
  We propose to model
  the fixation locations of the human eye when observing
  a still image by a Markovian point process in $\real^2$.  
  Our approach is data driven using k-means clustering of the 
  fixation locations to identify distinct salient regions of the image,
  which in turn correspond to the states of our Markov chain.  
  Bayes factors are computed as model selection criterion 
  to determine the number of clusters.
  Furthermore, we demonstrate that the behaviour of the human eye 
  differs from this model when colour information is removed
  from the given image.
\end{abstract}

\section{Introduction}


Ocular movement data has posed a particularly tough challenge to 
researchers and offers many potential insights into human
visual behaviour as well as many practical applications.
The contribution, if any, of colour information to
vision through saliency models and fixation location prediction 
has been heavily investigated; 
see \cite{bADDELEY2006}, \cite{FREY2008}, \cite{HO2012EYE}, \cite{HAMEL2014}.
Much research has gone into understanding ocular movement
from the rapidly jerking saccades to the relatively still 
fixations.  In this article, we will specifically focus on the eye's fixations
by modelling such a sequence of fixations as a point process
in $\real^2$.  
The distribution of fixations over a given image
is treated as a finite mixture model comprised of disjoint 
\textit{salient} regions, which correspond to the interesting bits
of the image; see \cite{MCLACHLAN2004}.
This set of salient regions is used as the state space of a Markov 
chain.  Each fixation is then an observation from the mixture 
component corresponding to the current state of the Markov chain.
Under this model, it is shown that the 
presence or absence
of colour information in the image drastically effects the 
behaviour of a given sequence of fixations.

\begin{figure}
  \includegraphics[width=\textwidth]{\PICDIR/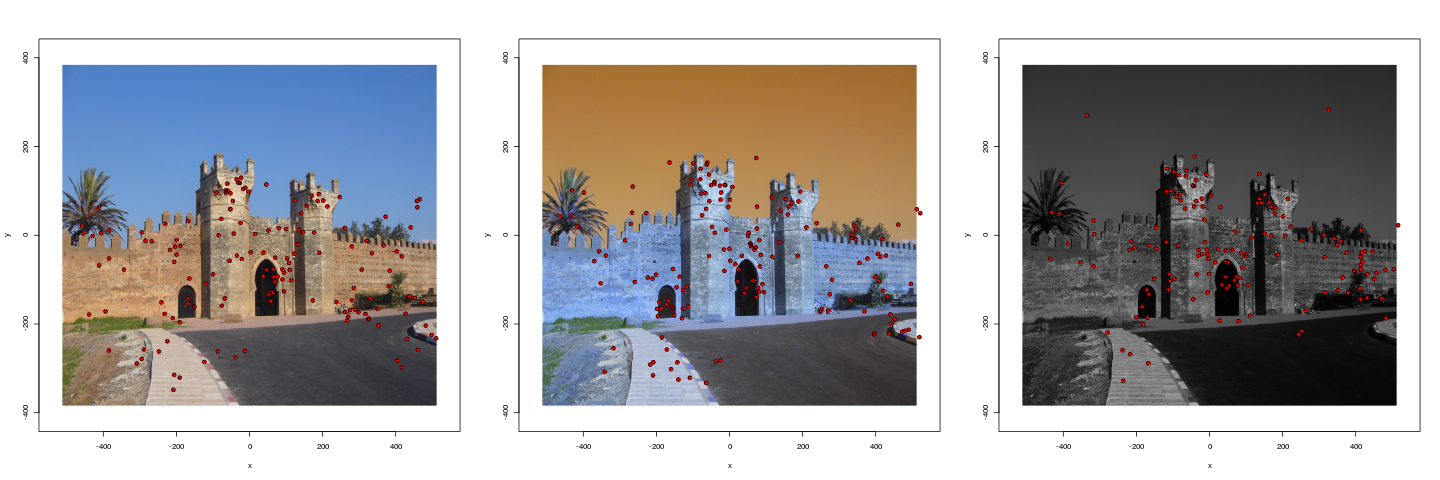}
  \caption{
    \label{fig:colourExample}
    An example of the three colour schemes, normal, abnormal, and
    grayscale, respectively, under analysis
    with plotted fixations.
  }
\end{figure}

The data under scrutiny comes for the study of 
\cite{HO2012EYE} who were interested as to whether the 
presence, absence, or modification of the colour of a
photograph affects how the eye moves when looking at a given
image.  There were three colour schemes in their study: normal
colours; abnormal colours; and grayscale.  Normal refers to the 
unmodified image.  Abnormal corresponds to swapping the red-green
and blue-yellow chrominance channels.  Grayscale corresponds to 
the complete removal of all colour information.  An example of the three
colour schemes with plotted fixations is displayed in 
Figure~\ref{fig:colourExample}.

The data was collected as follows.  Ten observers 
were selected for each of the three colour schemes totaling 30 
subjects in all.  Each subject was presented with 60 photographs
under a fixed colour scheme.  Each photo was displayed for five seconds,
and the position and duration of each fixation was recorded.  
An example of ten rows sampled from the data set are displayed in 
Table~\ref{tab:dataEx} whose entries from left to right 
are horizontal and vertical
position of the fixation, the duration in milliseconds, 
the fixation's sequence number,
the subject identifier, the colour scheme, 
the image number, and the orientation of
the image.  
A more detailed explanation of the 
data, the experiment, and the method of collection
can be found in \cite{HO2012EYE}.

Statistical analysis of static spatial point patterns and spatial
point processes has been used 
in this context; for an overview; see \cite{DIGGLE2003}, \cite{ILLIAN2008}.  
Modeling eye fixations as a spatial point process was previously
discussed in \cite{BARTHELME2013EYE} where an inhomogeneous Poison
process (IPP) was utilized.  The location dependent rate parameter of 
the IPP was determined by a measure of the saliency 
of each region of a given photograph.
Alternatively, \cite{KUMMERER2014EYE} apply deep neural
networks in order to identify salient regions and ultimately
to predict fixations.
But as is mentioned {in~\cite{HO2012EYE}},
``there is no computational saliency model that can predict an
observer's fixation location better than the model using 
fixations from other subjects.''
In light of that, we take a data driven approach to modelling
sequential fixation locations using nine of the
ten subjects to train our model and the tenth for validation.
Bayes factors are used as a model selection criterion; 
see \cite{GOOD1967} for the use of Bayes factors in the multinomial 
hypothesis setting,
and \cite{KASSRAFTERY1995} for a general overview of Bayes factors and
model selection.
With additional thought, 
our Markov states could ultimately be constructed from a 
saliency map of the image itself rather than from the data.

In this article, Section~\ref{sec:discrete} introduces a 
discrete time Markov model for the observed
sequences of ocular fixations.
The states are determined through $k$-means clustering where
cross validation is used to determine the optimal number of 
clusters.  A further investigation of alternative clustering methods,
a post-hoc look at the Markov transition probabilities, a closer analysis
of saccade lengths, and a display of the best and worst
scoring photographs under our model can be found in Sections~\ref{sec:cluster},
\ref{sec:trans}, \ref{sec:saccade}, and~\ref{sec:bestWorst}, respectively.
Section~\ref{sec:extend} discusses various potential
extensions to this model.
This includes Section~\ref{sec:continuous}, which
proposes reworking the discrete model as a continuous time Markov process
through a closer analysis of the fixations' durations, and 
Section~\ref{sec:imageDriven}, which discusses a saliency driven 
approach to model construction in contrast to our data driven method.
Lastly, Section~\ref{sec:conclusion} concludes with potential
applications.

\begin{table}
  \caption{
    \label{tab:dataEx}
    Ten randomly sampled entries from the data set.
  }
  \centering
  \fbox{
  \begin{tabular}{r@{.}lr@{.}lrrrrrr}
    \multicolumn{2}{c}{$X$} & \multicolumn{2}{c}{$Y$} & 
    Time & Fix. & Subj. & Colour & Image & Format\\
    \hline
      80&4&   74&8&  980&   10&  a5&  abnormal&    53& landscape\\
    -213&4&  111&6&  246&    6&  n5&    normal&    48& landscape\\
     499&9&  151&8&  241&   10&  a5&  abnormal&    11& landscape\\
     -32&7&  146&3&  150&    5&  a3&  abnormal&    51& landscape\\
    -256&3&  183&6&  135&   10&  n3&    normal&     4& landscape\\
     112&6&   86&0&  276&   12&  g7& grayscale&    47& landscape\\
     214&5& -133&9&  295&   10&  a4&  abnormal&    59& landscape\\
     409&3&    0&8&  225&   12&  a9&  abnormal&    30& landscape\\
     226&4& -343&8&  413&    3&  g2& grayscale&    44& landscape\\
    -111&0& -115&2&  157&    8& a10&  abnormal&    25& landscape
  \end{tabular}
  }
\end{table}

\section{Discrete time Markov model}
\label{sec:discrete}

\begin{figure}
  \begin{center}
  \includegraphics[width=\textwidth]{\PICDIR/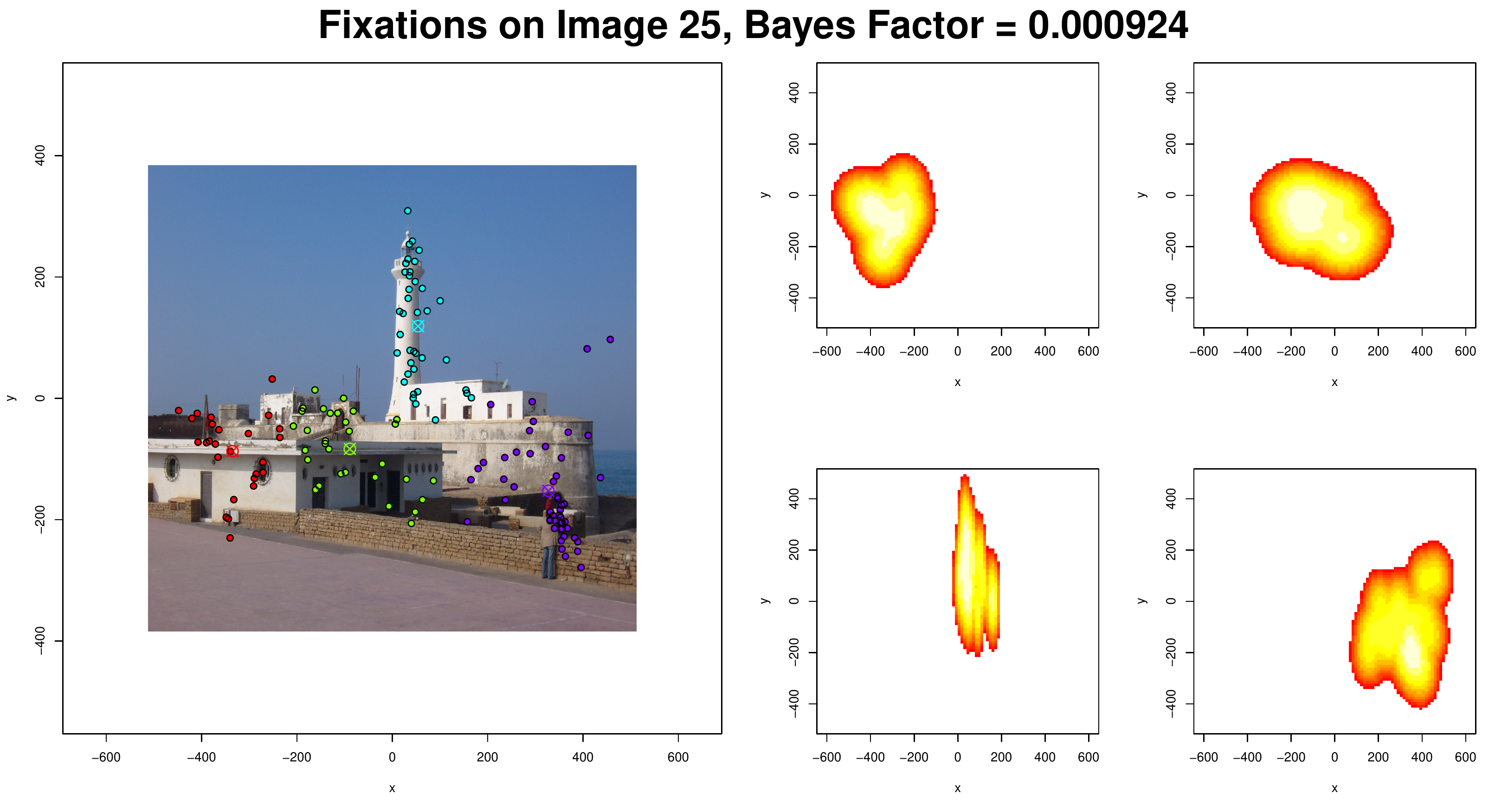}
  \end{center}
  \capt{\label{fig:clusterExample}
    {The four clusters of fixation locations for a given image, 
     and the corresponding kernel density estimates for the 
     fixation location model.}
  }
\end{figure}

Consider a sequence of fixation positions $X_1,\ldots,X_n\in\real^2$ 
as a point process in $\real^2$ and an associated sequence of 
states $S_1,\ldots,S_n\in\{1,\ldots,k\}$.
We will model this state sequence
as a Markov chain jumping between $k$ different clusters
corresponding to interesting parts of the photo.  The
fixation sequence will then be random observations conditioned on the 
current state of the Markov chain.
The model selection will decide between such models
for $k=1,\ldots,10$.
The $k=1$ case, our null model with which to compare the
others, is the naive model that the $X_t$ are
\iid draws from some underlying density $f(x)$. 
For $k\ge2$, we suppose a finite mixture model with $k$ constituent 
densities $f_1,\ldots,f_k$ 
corresponding to on which part of the image the eye is focusing.
In this model, the states evolve via 
a Markov chain with $X_t$ given by
an independent random draw from $f_{S_t}(x)$.

{These constituent densities were modelled empirically 
by clustering the fixation locations from nine of the ten subjects; 
the model was then tested on the tenth}.  
Cross-validation 
was performed across all {training} subjects to optimise this model.
Let $X_1,\ldots,X_n$
be the test sequence of {fixation locations} and $Y_{t}^{(j)}$ be fixation
$t$ of subject $j$ from the training set.
A Bayes factor 
was computed for each subject, and the results were averaged
into a final score for each picture.  The training fixation points
were clustered via $k$-means clustering with 10 random starts.
Other clustering methods are discussed in Section~\ref{sec:cluster}.
For each cluster, a two dimensional kernel density estimate
with Gaussian kernel
was computed. {An example of these clusters and density estimates 
can be seen in Figure~\ref{fig:clusterExample}.}  Each fixation
in the test set was assigned a cluster based on proximity to the
cluster center.  A $k$-nearest neighbours classifier was also 
implemented to assign clusters, but returned very similar results.

The observed initial states and transitions between states were 
treated as observations from a multinomial random variable
with a Dirichlet conjugate prior.
Specifically, the Markov initial, $\pi$, and transition 
probabilities, $p$,
were treated as Dirichlet random variables with the Jeffreys prior
and updated by the nine subjects in the training data.  
Let $c_i$ be the number of initial fixations $Y_1^{(j)}$ in state $i$,
and let $m_{i,i'}$ be the number of observed transitions
from $Y_{t-1}^{(j)}\in S_i$ to $Y_{t}^{(j)}\in S_{i'}$.
The posteriors are
\begin{align*}
  \pi &\dist \distDirich{ 0.5 + c_1,\ldots, 0.5 + c_k },\\
  p_{i,\cdot} &\dist \distDirich{
     0.5 + m_{i,1},\ldots, 0.5 + m_{i,k}
  }.
\end{align*}
Therefore, the Bayes factor is
$$
  BF 
  = \frac{\Prob{X_t}{Y_t,k=1}}{\Prob{X_t}{Y_t,k}}
  = \frac{\prod_{t=1}^n f(X_t)}{
    \xv_{\pi,p}\left\{ \pi_{s_1} f_{s_1}(X_1) \prod_{t=2}^n
    p_{s_{t-1},s_t} f_{s_t}(X_t) \right\}
  }
$$
where the expectation is taken with respect to the 
Dirichlet posterior.  In practice, this
value is approximated via Monte Carlo integration.

\begin{figure}
  \begin{center}
  \includegraphics[width=0.9\textwidth]{\PICDIR/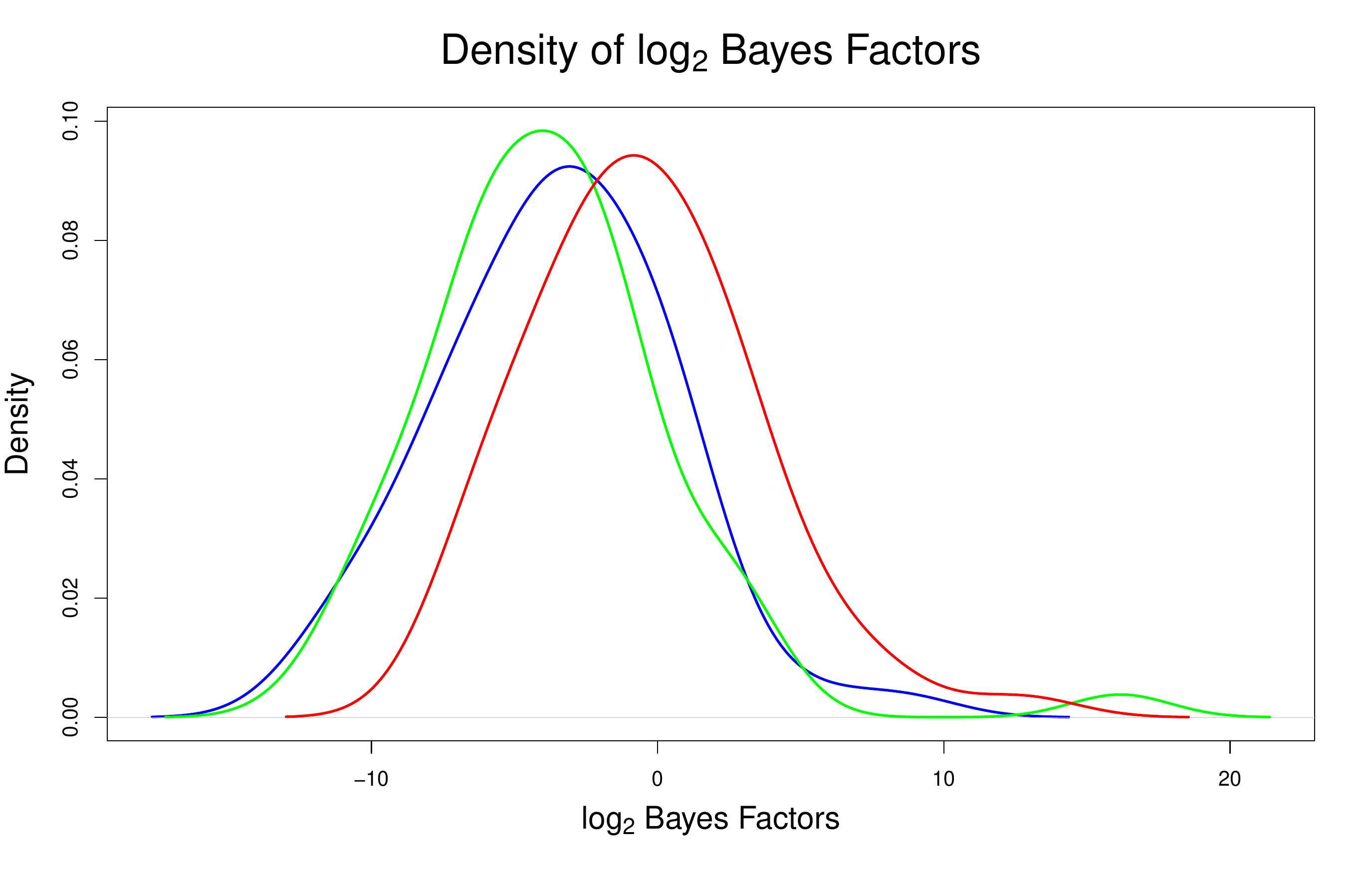}
  \end{center}
  \capt{\label{fig:clusterDensities}
    Density plots of the distribution of $\log_2$ Bayes factors 
    for images of each of the three colour schemes.  
    Blue is for normal, green is for abnormal, and red is for grayscale.
  }
\end{figure}

A plot depicting a kernel density estimates for the 
$\log_2$ Bayes factors of each colour scheme
is displayed in Figure~\ref{fig:clusterDensities}.
The red curve corresponding to the density of 
the grayscale Bayes factors appears shifted further 
to the right than the other two.  
Indeed, performing three paired t-tests 
results in the following 95\% confidence intervals
and p-values:
\begin{align*}
  &\texttt{Norm - Abno:}&~ &[-1.25,\phantom{-}1.71]&~
  \text{$p$-value} &= 0.76,\\ 
  &\texttt{Norm - Gray:}&~ &[-4.26,-1.77]&~ 
  \text{$p$-value} &= 9.6\times10^{-6},\\
  &\texttt{Abno - Gray:}&~ &[-4.61,-1.88]&~ 
  \text{$p$-value} &= 1.3\times10^{-5}.
\end{align*}
Consequently, the presence
of colour, whether normal or not, results in the majority of images
scoring a small Bayes factor, whereas the opposite is seen in
the grayscale setting.

\begin{figure}
  \begin{center}
  \includegraphics[width=0.6\textwidth]{\PICDIR/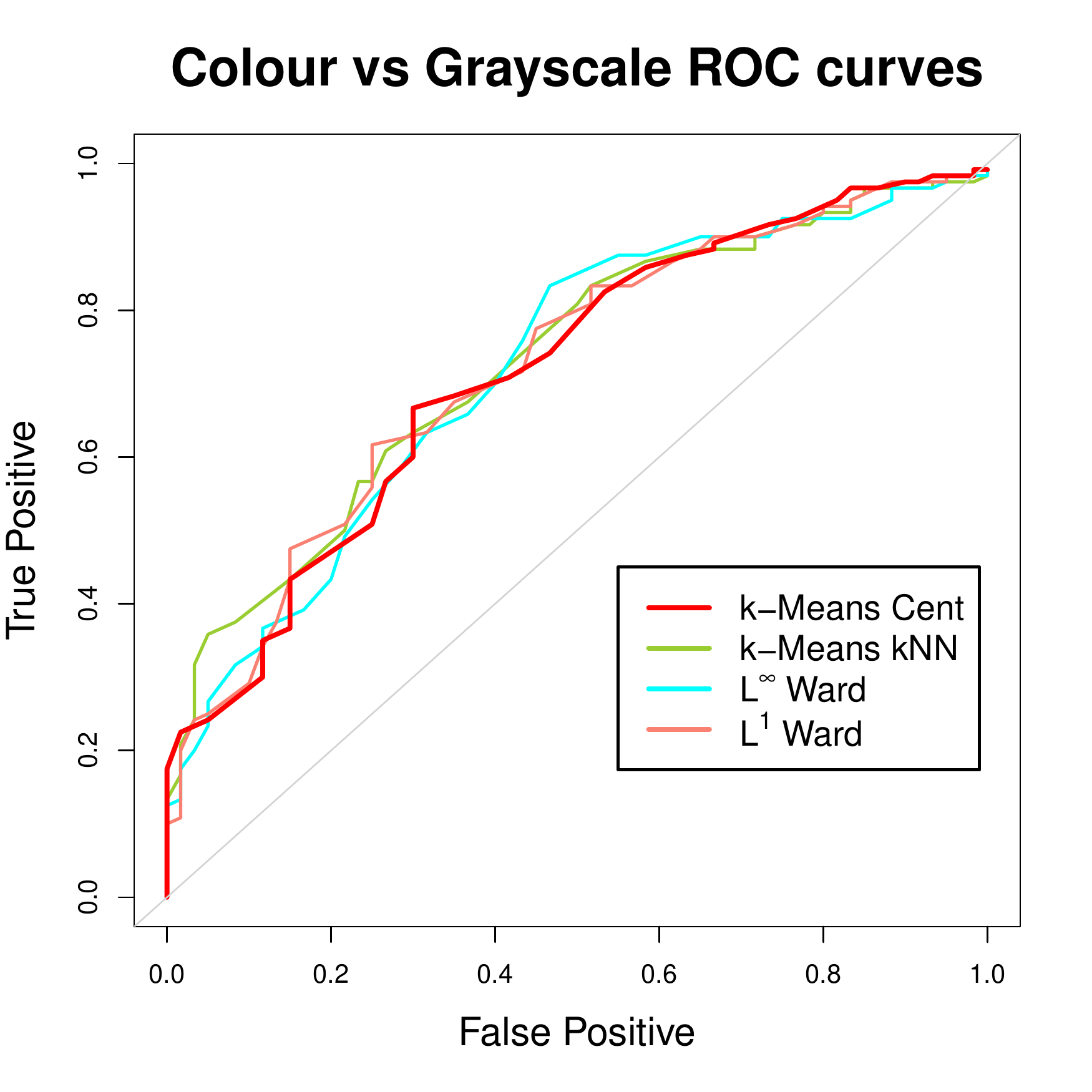}
  \end{center}
  \capt{ \label{fig:bfROC}
    ROC curves for thresholding on the Bayes Factor in order
    to determine whether the photo being viewed is in colour
    (true positive) or grayscale (false positive).  Four clustering
    methods are plotted with similar results.
  } 
\end{figure}

Furthermore, the Bayes factors for colour and grayscale images 
separate well enough that this model applied to observed sequences of
fixations can be used as a weak classifier as to whether or not 
the subjects are observing an image with colour information.
Indeed, over all of the 60 pictures
and 3 colour schemes, 14 normals, 13 abnormals,
but only 1 grayscale picture scored a Bayes factor $<0.01$.
The threshold that most separates this data set is $0.2$,
which correctly separates 66\% of the normals and 71\% of the
abnormals from the grayscale images.
Thresholding the Bayes factor as a classification criterion 
for whether or not the observed photograph has colour, normal
or abnormal,
results in the ROC curves of Figure~\ref{fig:bfROC}. 
The ROC curves consider clustering with $k$-means where inclusion is
based on either proximity to the cluster centre or $k$-nearest
neighbours.  Two hierarchical clustering methods are also included,
which will be discussed more in Section~\ref{sec:cluster}.
Here,
`true positive' refers to the percentage of coloured photos 
with Bayes factor below the threshold and `false positive' for 
the percentage of grayscale photos below the threshold. 

Ultimately,
a sequence of ocular fixations in the
grayscale case is better modelled as a collection of independent
random draws.  In contrast, the coloured cases are better modelled 
as if jumps between interesting regions of the image occur 
in a Markovian fashion.
This suggests that the absence of colour can 
make it more difficult for subjects to identify
and scan through 
interesting parts of an image.  


\subsection{Clustering methods}
\label{sec:cluster}

\begin{figure}
  \begin{center}
  \includegraphics[width=\textwidth]{\PICDIR/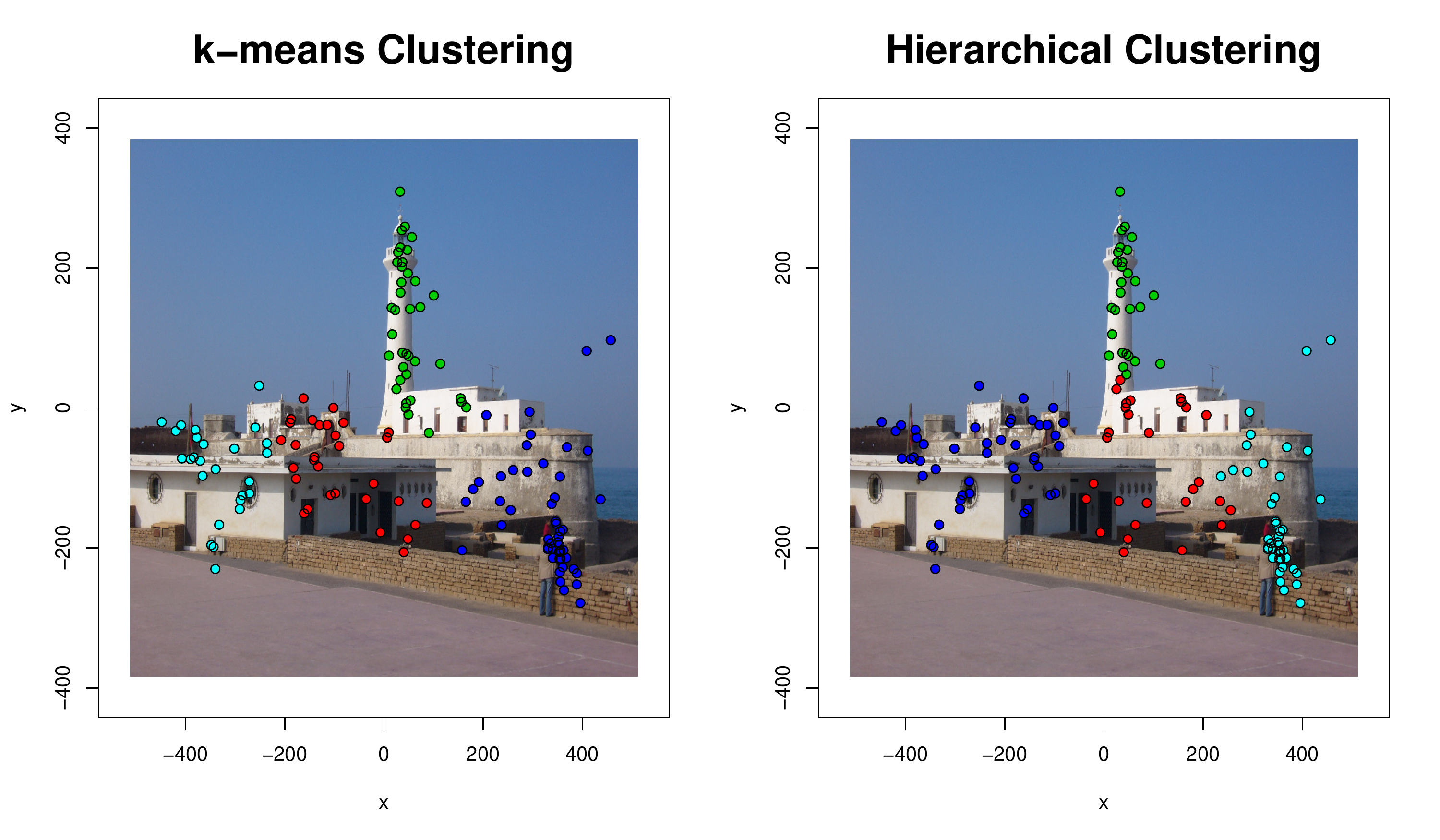}
  \end{center}
  \capt{ \label{fig:compClust}
    A comparison of $k$-means and hierarchical clustering.
  }
\end{figure}

The use of $k$-means clustering with Euclidean distance 
puts an heavy assumption on 
our model.  Specifically, this approach partitions a
photograph into Voronoi cells, which are by design all convex polygons.  
This approach strives to construct spherical and similarly sized
clusters specifically removing the possibility of non-convex or 
nested clusters.  In light of this, a variety of agglomerative
hierarchical clustering methods were also tested.
For example, Figure~\ref{fig:compClust} depicts $k$-means clustering
of the fixations of image~25 using Euclidean distance 
on the left and complete-linkage hierarchical clustering 
using the Manhattan or $L^1$ distance of those same fixations 
on the right resulting in different clusters forming.

This ``bottom-up'' hierarchical clustering begins with each fixation
occupying its own cluster.  The method iteratively combines clusters 
based on 
a combining criterion and an underlying metric.  In our analysis,
the chosen metrics to test were the Manhattan or $L^1$, the 
Euclidean or $L^2$, and the maximum or $L^\infty$ distances.
The linkage methods chosen were Ward's minimum variance method,
\cite{WARD1963} and \cite{MURTAGH2014}, complete linkage clustering, 
and unweighted pair group method with arithmetic mean (UPGMA), 
\cite{SOKAL1958}.
See Section~14.3 of \cite{HASTIE2005} or 
Section~8.5 of \cite{LEGENDRE2012} for an overview of such
methods.

Of the various combinations of such metrics and linkage criteria,
none performed noticeably better than $k$-Means, and many
combinations performed worse.  The ROC curves in 
Figure~\ref{fig:bfROC} include two for Ward's method with
$L^\infty$ and $L^1$ distances.  As the clusters formed 
via hierarchical clustering need not be convex, $k$-nearest
neighbours was used to determine to which cluster a given 
fixation from the testing set belonged with a variety of $k$
tested. 

\subsection{Analysis of transitions}
\label{sec:trans}

As a specific example, we will consider image 25 under the 
normal colour scheme, which is displayed in 
Figure~\ref{fig:clusterExample}.  Running the above analysis
yielded a decisively strong Bayes factor of 0.0009 in favour 
of the $k=4$ model over the $k=1$ model.  The states are 
coloured red, green, cyan, purple moving left to right 
across the image.

For each of the 
ten subjects, the maximum likelihood estimate of the 
initial probabilities and transition matrix from the
Dirichlet posteriors were averaged into the following:
\begin{align*}
  \pi = &\begin{pmatrix} \,\,0.05& 0.45& 0.13& 0.37\, \end{pmatrix}, \\
  p = &\begin{pmatrix}
    0.51& 0.29& 0.14& 0.06\\
    0.30& 0.26& 0.24& 0.20\\
    0.07& 0.18& 0.58& 0.18\\
    0.05& 0.11& 0.13& 0.70
  \end{pmatrix}.
\end{align*}
Here, states 1, 3, and 4 fall into the often seen pattern of 
having probability higher than 50\% of remaining in the same 
state and of having other transition probabilities that roughly
decrease as the distance between clusters increases.

\subsection{Analysis of saccade length}
\label{sec:saccade}

We now investigate the saccades, which are the distances
between successive fixations.  We make use of the 
Euclidean distance of the fixations on a plane 
for the following analysis.  In \cite{HO2012EYE},
saccades are instead measured with an angular metric.

\begin{figure}
  \begin{center}
  \includegraphics[width=0.47\textwidth]{\PICDIR/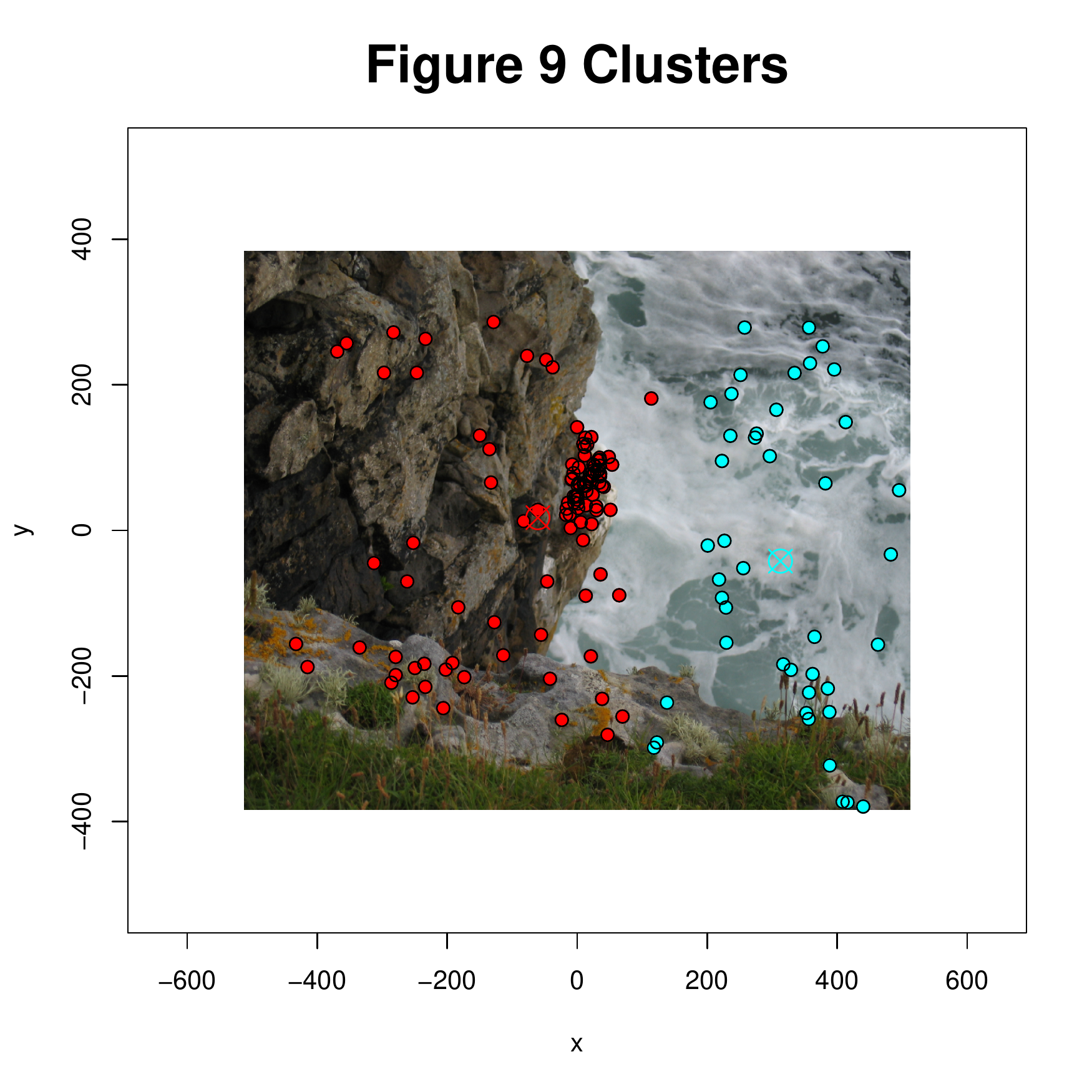}
  \includegraphics[width=0.47\textwidth]{\PICDIR/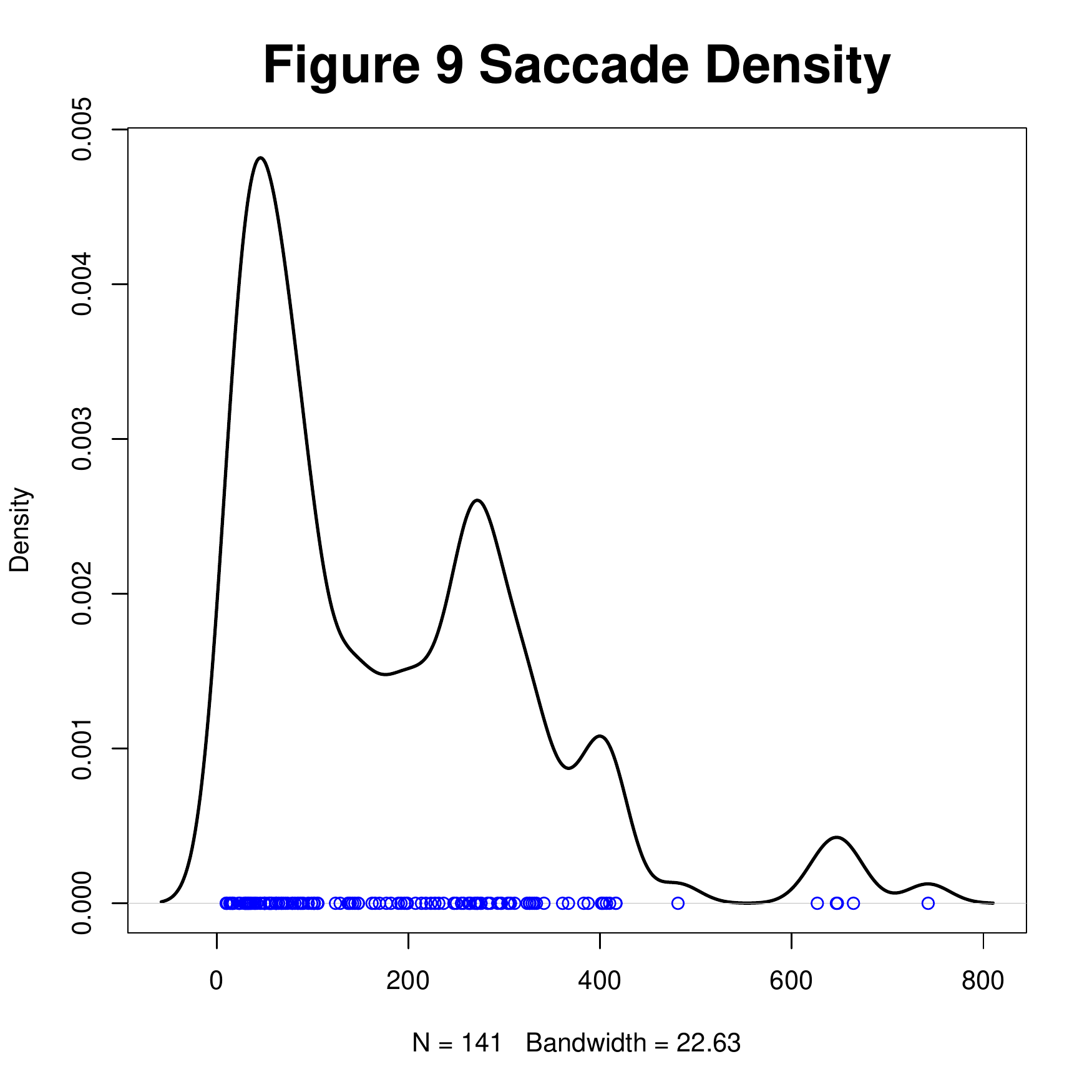}
  \end{center}
  \capt{\label{fig:sacDen}
    On the left, two clusters of fixations.  On the right,
    a kernel density plot of the saccade lengths.
  }
\end{figure}

For images that strongly fit the above Markov model,
the saccades
follow a natural mixture model.  For example, if the Markov
point process is supported on two states, then the eye 
can choose to stay in the same state (i.e. a short saccade)
or transition to the other state (i.e. a long saccade).  
This behaviour is readily evident in Figure~\ref{fig:sacDen},
which displays the two clusters of fixations for image~9 on the
left and a kernel density estimate for the distribution of
the saccade lengths on the right.  The KDE was computed with
a Gaussian kernel using the Sheather-Jones method of bandwidth
selection, \cite{SHEATHER1991}.

The saccades, however, contain further information than just reinforcing 
the Markovian structure of the data.  As reported in 
\cite{HO2012EYE}, a Kolmogorov-Smirnov (KS) test comparing the 
empirical distributions of the saccades across all images between 
the normal and 
abnormal colour schemes yields a significant p-value.
Under our use of Euclidean distance, the following are the
three p-values for the three KS tests:
\begin{align*}
  \texttt{Norm - Abno:}&~ p = 0.0163,& 
  \texttt{Norm - Gray:}&~ p = 0.394,&
  \texttt{Abno - Gray:}&~ p = 0.116
\end{align*}

Going further, we normalise the saccades of each of the 30 
subjects by subtracting the subject's sample mean and dividing by the
sample standard deviation.  The goal is to remove inter-subject
variability from the data in order to focus only on the 
between colour scheme variation.  The following p-values are
from the KS test on the normalised saccades.
\begin{align*}
  \texttt{Norm - Abno:}&~ p = 0.0264,& 
  \texttt{Norm - Gray:}&~ p = 0.0471,&
  \texttt{Abno - Gray:}&~ p = 0.788
\end{align*}

The main point of interest is the significant difference 
between the saccades under the normal and abnormal colour 
schemes.  In the previously detailed Markov model, there is
no discernible difference between these two settings.
Hence, a further understanding of the differences in the
saccades could enhance the Markov model.

\subsection{The best and the worst}
\label{sec:bestWorst}

\begin{figure}
  \centering
  \includegraphics[width=\textwidth]{\PICDIR/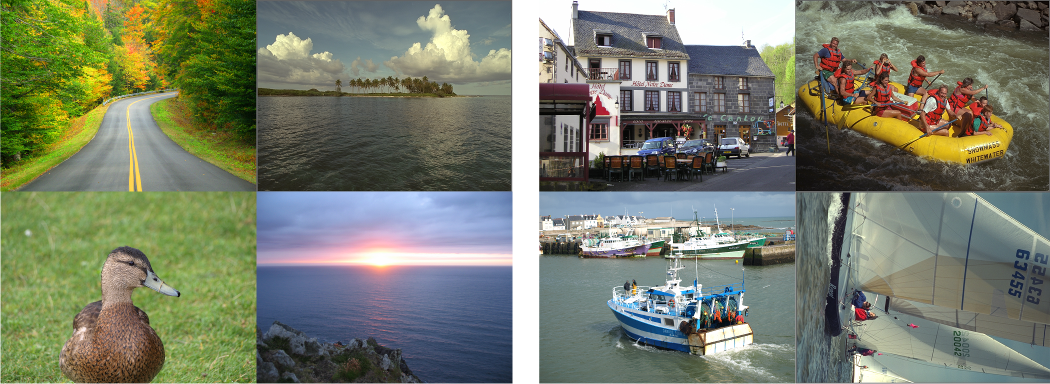}
  \vspace{0.01in}
 
  \includegraphics[width=\textwidth]{\PICDIR/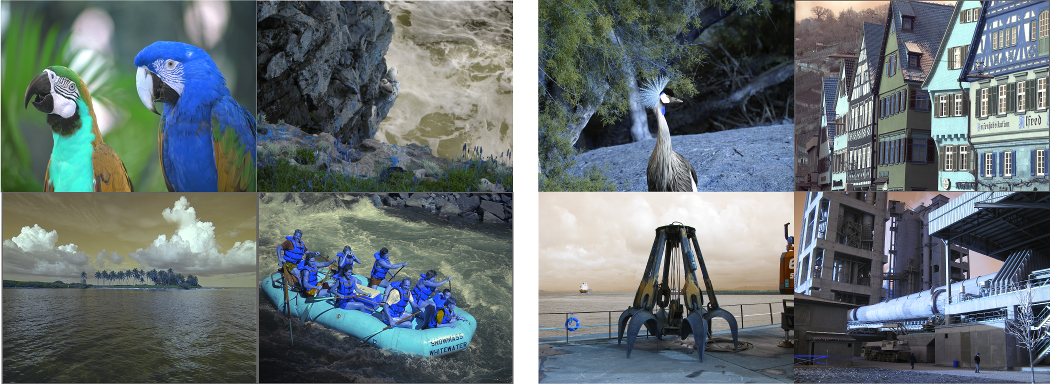}
  \vspace{0.01in}
 
  \includegraphics[width=\textwidth]{\PICDIR/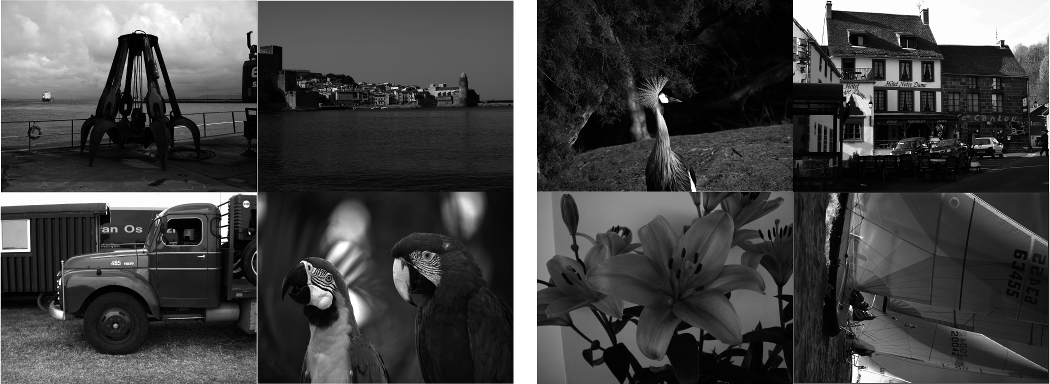}
  \capt{ \label{fig:bestWorst}
    The four worst scoring pictures on the left and the 
    four best scoring pictures on the right for the three
    colour schemes.
  } 
\end{figure}

Using the strongest Bayes factor for each photograph, we
can rank the photos in order of which best fits the Markov
model for $k\ge2$ clusters.  The four best and worst photos
for each colour scheme are depicted in Figure~\ref{fig:bestWorst}.
The normal coloured images are reasonably partitioned as
the worst scoring images contain a singular point of 
focus whereas the best scorers contain multiple objects on 
which to fixate such as text and people.
The abnormal setting
gives similar results barring the photo of eight people in 
a raft, which scores strongly under normal colours but produced
the single worst score under the abnormals.  
Presumably, the abnormal colour scheme most
disorients the brain when it is applied to objects with a narrow
range of expected colours such as human faces, which 
are generally not blue.

\section{Extensions}
\label{sec:extend}

In the following subsections, extensions to the Markov
model, which were not fully investigated, are detailed.
Each, if incorporated correctly, has the potential to 
add a great deal of insight into our model.

\subsection{Continuous time setting}
\label{sec:continuous}

Perhaps the most blatant omission from the previously
described model is the time spent at each fixation.
A clever model of the fixation durations would allow 
for the construction of a continuous time Markov process
adding more depth to the model.
Evidence suggests that the colour scheme, the fixation
location, and the amount of time spent staring at the
photograph all contribute to the fixation duration.

First, there are notable differences in the fixation durations
under the three colour schemes.  In the article 
of \cite{HO2012EYE}, 
Kolmogorov-Smirnov tests were run between each pair of
empirical distributions for the overall fixations durations.
They report no significant difference between the 
normal and abnormal settings, but report high significance
between the grayscale and each of the coloured settings.
Furthermore, the distribution of the fixation durations is
shown to be non-stationary in time.  That is, the distribution
of initial durations differs from the distribution of later
durations.

From our own investigations, we ask 
how fixation duration
correlates with what is being observed? 
To answer this, 
a two dimensional density estimate
{based on the fixation durations of nine of the ten subjects} was used as   
a {data-driven} measure of how interesting a region of a 
given photograph is.
Computing the correlation
between the density estimate at each fixation point and the
duration of tenth subject's {fixation at} that point gave the 
following nontrivial value 
of 0.1838.  The associated 95\% confidence interval
is {$[0.1721, 0.1954]$}. {This demonstrates that the fixation 
durations of previous subjects can indeed provide useful
predictive information about the fixation times for future subjects.}

\begin{figure}
  \begin{center}
  \includegraphics[width=\textwidth]{\PICDIR/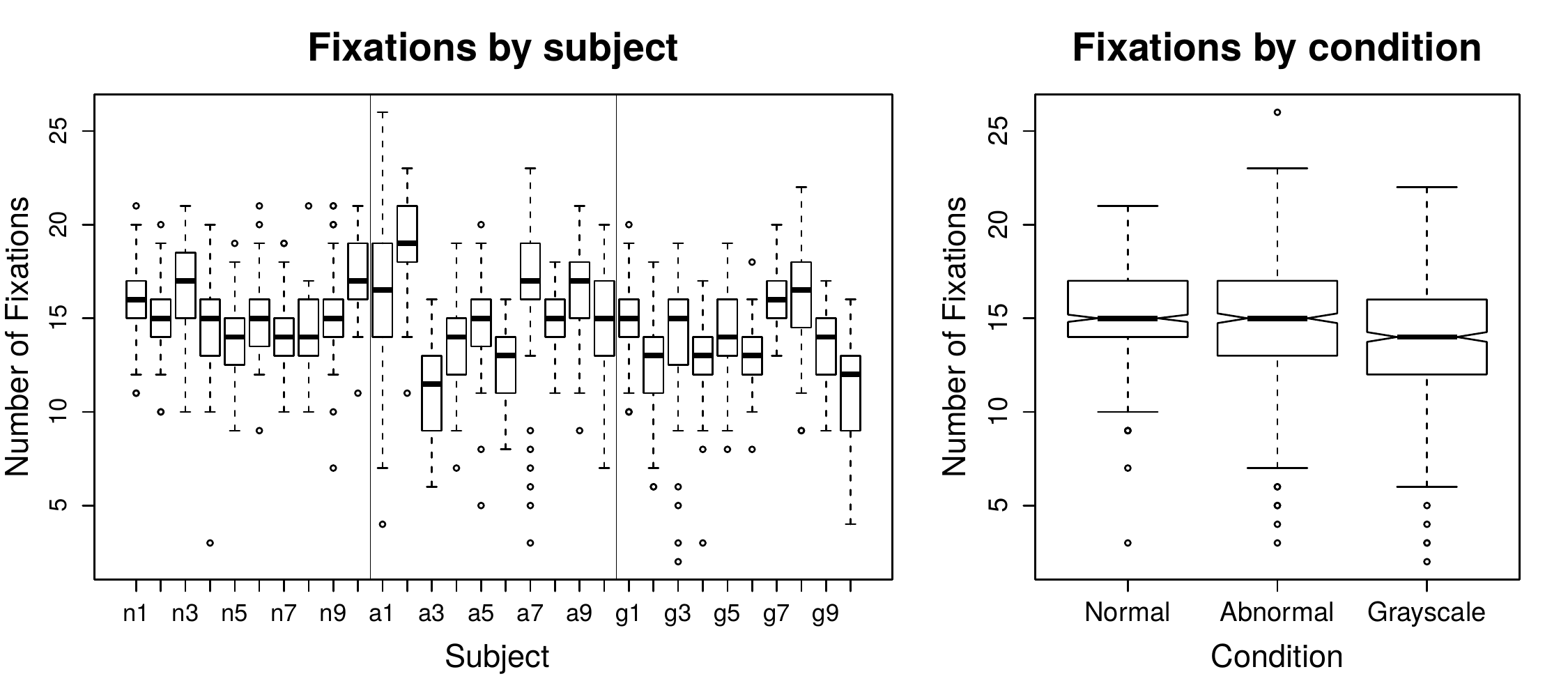}
  \end{center}
  \capt{ \label{fig:fixBoxPlot} 
    Box plots of the total number of fixations partitioned 
    based on subject and condition.
  }
\end{figure}

Due to the finite time each picture was displayed,
one can {also} analyse fixation duration by considering its inverse 
relationship with respect to the total number of fixations 
per subject per image.  The number of fixations per subject
per image {were} tallied and partitioned into three sets for
normal, abnormal, and grayscale colour.  The box 
plot in Figure~\ref{fig:fixBoxPlot} depicts the spread of the three sets
of data. It is visibly evident that the normal colour scheme
results in a much tighter grouping than the others.  Due to the
discreteness of the data, the equality of means was tested with
the Kruskal-Wallis test, which returned a $p$-value less than 
$10^{-15}$.  Comparing the three pairs via the Mann-Whitney
tests resulted in $p$-values:
\begin{align*}
  \texttt{Norm - Abno:}&~ p = 0.657,& 
  \texttt{Norm - Gray:}&~ p < 10^{-15},&
  \texttt{Abno - Gray:}&~ p = 4\times10^{-11} 
\end{align*}
This demonstrates that on average there were fewer fixations
made on the grayscale images. {In agreement with our results above, this suggests that} the lack of colour
information resulted in longer fixations and fewer in total 
as the eye took more time to understand what it was viewing.

\subsection{Image driven analysis}
\label{sec:imageDriven}

Rather than relying on the data to construct clusters,
one could look to the image itself.  First and foremost, 
there has been a considerable amount of research on
saliency maps such as \cite{KUMMERER2014EYE} who take
inspiration from computer vision and use deep neural networks to
determine the eye-catching parts of an image.
Beyond computer vision and saliency, there is a vast literature
on image segmentation.  Ultimately a multifaceted model could
be used to partition an image into distinct regions of interest.
Such a model may include image gradient information to identify
sharp edges and corners that draw the eye as well as
specific object recognition to mimic 
the human eye's affinity for identifying human faces or written
words, which will demand more fixations.

\section{Summary and Discussion}
\label{sec:conclusion}

Ocular fixation data yields both a challenging and useful 
analysis. We demonstrated that in the presence of colour
information, a sequence of fixations from a human eye can
be modeled as a Markov point process.  Furthermore, this 
model breaks down when such colour information is removed
from the image.  Given that we believe this model, 
a thorough analysis of the Dirichlet posteriors 
could yield interesting insight as to how different
photographs are treated by the human eye.

This model has the potential to lead to future applications 
such as a 
passive diagnostic test for sudden loss of colour vision.
That is, even with the 60 miscellaneous photographs of the
given data set, it is still possible to roughly classify 
which observers are looking at colour images and which are not.
With a carefully constructed set of colour photographs to elicit 
Markovian eye movements, one could then use collected
data from healthy eyes to train a classifier to determine
whether a patient can see colour or not with no other 
active participation from the patient besides staring at
the set of diagnostic photographs.

Ultimately, there is much room for further analysis.  The 
saccade lengths hint at differences between how the eye 
handles normal versus abnormal colour schemes.  Successful
integration of the the temporal information will lead to a 
more comprehensive model.  Attempting to construct such a 
model from image information rather than fixation data
could offer insight into saliency maps and lead to more
complex and interesting Markov states than the convex polygons
from the $k$-means approach.

\section{Acknowledgments}

The authors would like to acknowledge the Young Statisticians 
Section and Research Section of the Royal Statistical Society 
and the contest sponsor, Select Statistics,
for constructing a very fun and intellectually exciting 
Statistical Analytics Challenge.


\bibliographystyle{rss}
\bibliography{kasharticle,kashbook}

\end{document}